\documentclass[a4paper]{jpconf}
\usepackage{graphicx,amsbsy}
\newcommand{\apjref}[3]{#3 \textit{Astrophys.\ J} \textbf{#1}, {#2}}
\newcommand{\naturef}[3]{#3 \textit{Nature} \textbf{#1}, {#2}}
\newcommand{\prdref}[3]{#3 \textit{Phys.\ Rev.}\ \textbf{D#1}, {#2}}
\newcommand{\prlref}[3]{#3 \textit{Phys.\ Rev.\ Lett.}\ \textbf{#1}, {#2}}
\newcommand{\cqgref}[3]{#3 \textit{Class.\ Quant.\ Grav.}\ \textbf{#1}, {#2}}
\newcommand{\mc}[1]{\mathcal{#1}}
\newcommand{\text}[1]{\mathrm{#1}}

\newcommand{\ev}[1]{\left\langle#1\right\rangle}

\newcommand{\eqref}[1]{(\ref{#1})}
\newcommand{\dcc}{P1200051-v4}
\begin{document}
\title{Treatment of Calibration Uncertainty in Multi-Baseline
  Cross-Correlation Searches for Gravitational Waves}

\author{John T.\ Whelan}

\address{Center for Computational Relativity and Gravitation
  and School of Mathematical Sciences, Rochester Institute of Technology,
  85 Lomb Memorial Drive, Rochester NY 14623, USA}

\ead{john.whelan@astro.rit.edu}

\author{Emma L.\ Robinson}

\address{Max-Planck-Institut f\"{u}r Gravitationsphysik
  (Albert-Einstein-Institut), D-14476 Potsdam, Germany}

\author{Joseph D.\ Romano}

\address{Center for Gravitational-Wave Astronomy and Department of
  Physics and Astronomy, The University of Texas at Brownsville, 80
  Fort Brown, Brownsville TX 78520, USA}

\author{Eric H.\ Thrane}

\address{Department of Physics, University of Minnesota, Minneapolis,
  MN, USA}

\begin{abstract}
  Uncertainty in the calibration of gravitational-wave (GW)
  detector data leads to systematic errors which must be accounted for
  in setting limits on the strength of GW signals. When
  cross-correlation measurements are made using data from a pair of
  instruments, as in searches for a stochastic GW background, the
  calibration uncertainties of the individual instruments can be
  combined into an uncertainty associated with the pair. With the
  advent of multi-baseline GW observation (e.g., networks consisting
  of multiple detectors such as the LIGO observatories and Virgo), a
  more sophisticated treatment is called for. We describe how the
  correlations between calibration factors associated with different
  pairs can be taken into account by marginalizing over the
  uncertainty associated with each instrument.
\end{abstract}

\section{Calibration Uncertainty with One Baseline}

Consider an experiment to measure a physical quantity ${\mu}$ (e.g.,
the stochastic GW background strength
$\Omega_{\text{gw}}(f)$~\cite{Nelson92,AllenRomano}).  An optimal
combination ${x}$ of cross-correlation measurements provides a point
estimate of ${\mu}$ with error bar ${\sigma}$.  Given a likelihood
function $p({x}|{\mu})$ and a prior $p(\mu)$, one can use Bayes's 
Theorem to construct the posterior
\begin{equation}
\label{e:bayes}
p({\mu}|{x})
= \frac{p({x}|{\mu})p({\mu})}{p({x})}
= \frac{p({x}|{\mu})p({\mu})}
{\int d{\mu}\,p({x}|{\mu})\,p({\mu})}\,.
\end{equation}

Due to calibration uncertainties in each of the instruments that 
make up the baseline for the cross-correlation, ${x}$ is an estimator
not of ${\mu}$, but of ${\lambda}{\mu}$, where
${\lambda}$ is an unknown calibration factor (for the baseline)
described by an
uncertainty ${\varepsilon}$.  Thus the likelihood function also depends on
the calibration factor ${\lambda}$
\begin{equation}
p({x}|{\mu},{\lambda})
= \frac{1}{{\sigma}\sqrt{2\pi}}
\,\exp
\left[
  -\frac{({x}-{\lambda}{\mu})^2}{2{\sigma}^2}
\right]
\ ,
\end{equation}
so the posterior \eqref{e:bayes} is now constructed from 
the {\em marginalized} likelihood
\begin{equation}
p({x}|{\mu})
= \int d{\lambda}
\,p({x}|{\mu},{\lambda})p({\lambda})
\ .
\end{equation}
If we assume a Gaussian distribution for $\lambda$,
\begin{equation}
p({\lambda})
= \frac{1}{{\varepsilon}\sqrt{2\pi}}
\,\exp\left[-\frac{({\lambda}-1)^2}{2{\varepsilon}^2}\right]
\ ,
\end{equation}
then we can do the marginalization analytically if 
the range of
$\lambda$ values is taken to be $(-\infty,\infty)$.%
\footnote{Since $\lambda$ is an amplitude calibration factor,
it should take on only positive values.
But for small $\varepsilon$---e.g., of order 0.10 to 
0.20---integrating over $\lambda$ from $0$ to $\infty$
is numerically the same as integrating
over $\lambda$ from $-\infty$ to $+\infty$.}
This leads to
\begin{equation}
p({x}|{\mu})
= 
\frac{1}{\sqrt{2\pi (\sigma^2+{\varepsilon}^2{\mu}^2)}}
\ \exp
\left[
  -\frac{1}{2}
  \frac{(x-{\mu})^2}{(\sigma^2+{\varepsilon}^2{\mu}^2)}
\right]\,.
\label{e:single}
\end{equation}
This is the method used in stochastic GW searches with two LIGO sites,
e.g.,~\cite{S4stoch,S5stoch}.

\begin{figure}
  \begin{center}
     \includegraphics[width=0.45\columnwidth]{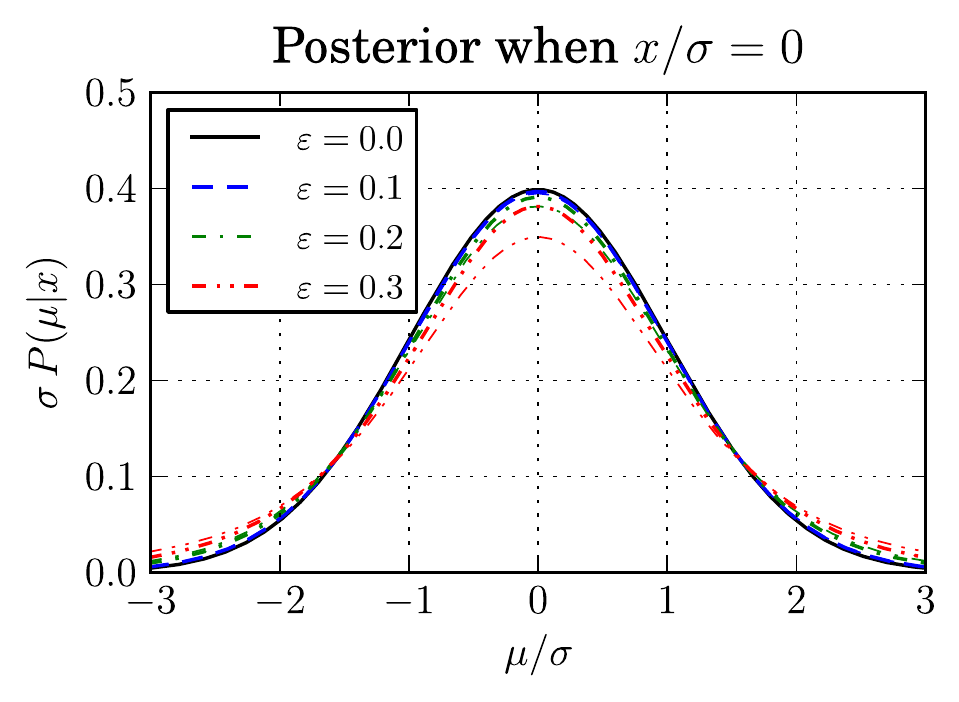}
     \includegraphics[width=0.45\columnwidth]{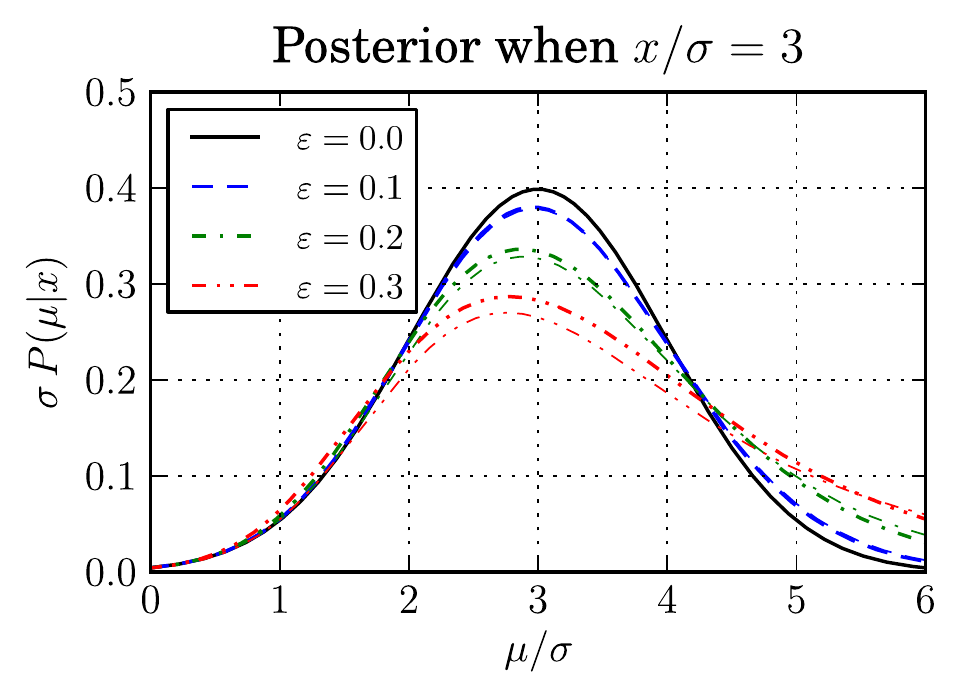}
   \end{center}
   \caption{Effects of marginalizing analytically over a single
     calibration factor for two different values of the measured
     cross-correlation, $x=0$ and $x=3\sigma$.  The thick line is a
     numerical marginalization with a log-normal prior on ${\lambda}$;
     the thin line is an analytic marginalization with a Gaussian
     prior on $\lambda$.}
   \label{fig:calmarg}
\end{figure}

A more physically-motivated prior, which explicitly
takes into account that $\lambda$ is multiplicative and 
takes on only positive values, is a log-normal distribution
\begin{equation}
p({\lambda})
= \frac{1}{{\lambda}{\varepsilon}\sqrt{2\pi}}
\,\exp\left[-\frac{(\ln{\lambda})^2}{2{\varepsilon}^2}\right]
\quad{\rm or}\quad
p({\Lambda})
= \frac{1}{{\varepsilon}\sqrt{2\pi}}
\,\exp\left[-\frac{{\Lambda}^2}{2{\varepsilon}^2}\right]
\quad
\hbox{where\ }
{\Lambda}=\ln{\lambda}\,.
\end{equation}
This was the approach taken in the stochastic GW search using LIGO and
ALLEGRO~\cite{S4L1A1}, but has the drawback of requiring numerical
integration over ${\Lambda}$ because
\begin{equation}
p({x}|{\mu},{\Lambda})
= \frac{1}{{\sigma}\sqrt{2\pi}}
\,\exp
\left[
  -\frac{({x}-e^{{\Lambda}}{\mu})^2}{2{\sigma}^2}
\right]
\end{equation}
gives a factor which is not Gaussian in ${\Lambda}$.
Figure~\ref{fig:calmarg} compares the posterior
distributions for the two different choices of priors
for different measured values of $x$ and different values of $\varepsilon$.

\section{Calibration Uncertainty with Multiple Baselines}

With more than two instruments, there are multiple baselines and
multiple calibration uncertainties to marginalize over.  For instance,
the stochastic background search using initial LIGO and Virgo
data~\cite{ligovirgo,S5LV} involved 4 different instruments
$I\in\{\text{H1},\text{H2},\text{L1},\text{V1}\}$ and 5 different
baselines
$\alpha\in\{\text{H1L1},\text{H1V1},\text{H2L1},\text{H2V1},\text{L1V1}\}$.%

Since the cross-correlation measurements for different baselines
involve different calibration factors, all of the baselines cannot be
optimally combined before marginalizing over calibration.  Instead,
all of the measurements for a baseline $\alpha$ can be combined into a
single point estimate ${x_\alpha}$ with error bar ${\sigma_\alpha}$.
Each baseline has unknown calibration factor ${\lambda_\alpha}$.
Since the statistical errors for the different baselines are
independent \cite{lazz}, the likelihood is the product
\begin{equation}
p(\mathbf{x}|{\mu},\boldsymbol{\lambda})
= \prod_\alpha\left\{\frac{1}{{\sigma_\alpha}\sqrt{2\pi}}
\,\exp
\left[
  -\frac{({x_\alpha}-{\lambda_\alpha}{\mu})^2}
  {2{\sigma_\alpha}^2}
\right]\right\}
\ ,
\end{equation}
where $\mathbf{x}\equiv\{x_\alpha\}$ and
$\boldsymbol{\lambda}\equiv\{\lambda_\alpha\}$.
The calibration factor ${\lambda_\alpha}$ for each baseline is
${\lambda_{IJ}}={\xi_I}{\xi_J}$, and is determined by the
per-instrument calibration factors $\boldsymbol{\xi}\equiv\{\xi_I\}$.
If each instrument's calibration has an underlying uncertainty
${\delta_I}$, the per-baseline calibration factors
$\boldsymbol{\lambda}$ have the following means, variances and
covariances:
\begin{equation}
\ev{{\lambda_{IJ}}} = 1\,,
\quad
\ev{{\lambda_{IJ}\lambda_{IJ}}}
= 1 + {\delta_I}^2 + {\delta_J}^2
+ \mc{O}({\delta}^4);
\quad
\ev{{\lambda_{IJ}\lambda_{JK}}}
= 1 + {\delta_J}^2
+ \mc{O}({\delta}^4)
\quad \hbox{if $I\ne K$}
\ .
\end{equation}

\subsection{Per-Baseline Calibration Marginalization}

One approach is to marginalize over the per-baseline calibration
factors assuming a multivariate Gaussian prior $p(\boldsymbol{\lambda})$.  
This has the advantage that the marginalization integral
\begin{equation}
p(\mathbf{x}|{\mu})
= \int d{\boldsymbol{\lambda}}\>
p(\mathbf{x}|{\mu},\boldsymbol{\lambda})
\,p(\boldsymbol{\lambda})
\end{equation}
can be done analytically if the integrals over the per-baseline
calibration factors $\boldsymbol{\lambda}$ are taken over
$(-\infty,\infty)$.  However, the relationship
${\lambda_{IJ}}={\xi_I}{\xi_J}$ implies that
\begin{equation}
  \label{e:lambdaident}
{\lambda_{IJ}}{\lambda_{KL}}
- {\lambda_{IK}}{\lambda_{JL}} = 0
\ .
\end{equation}
For a multivariate Gaussian prior on $\boldsymbol{\lambda}$,
this relation is true {\em only as an expectation value}, 
not as an identity for all values of $\lambda_{IJ}$, $\lambda_{KL}$,
etc.

\subsection{Per-Instrument Calibration Marginalization}

An alternative approach which enforces identities such as 
\eqref{e:lambdaident} is
to set a prior which is the product of independent priors on each
per-instrument calibration factor ${\xi_I}$ or equivalently on
${\Xi_I}=\ln{\xi_I}$.  Similarly defining the log-calibration
factor for a baseline ${\Lambda_\alpha}=\ln{\lambda_\alpha}$,
we have
\begin{equation}
{\Lambda_{IJ}} = \ln{\lambda_{IJ}}
= \ln({\xi_I}{\xi_J})
= {\Xi_I} + {\Xi_J}
\ .
\end{equation}
The likelihood is
\begin{equation}
p(\mathbf{x}|{\mu},\boldsymbol{\Xi})
= \prod_\alpha\left\{
\frac{1}{{\sigma_\alpha}\sqrt{2\pi}}\,
\exp
\left[
  -
  \frac{({x_\alpha}-e^{{\Lambda_\alpha}}{\mu})^2}
  {2{\sigma_\alpha}^2}
\right]
\right\}
\end{equation}
and the marginalized likelihood is
\begin{equation}
p(\mathbf{x}|{\mu})
= \int d\boldsymbol{{\Xi}}\>
p(\mathbf{x}|{\mu},\boldsymbol{\Xi})
\,p(\boldsymbol{\Xi})
\ .
\end{equation}
An obvious prior is log-normal on $\xi_I$, i.e., Gaussian on $\Xi_I$:
\begin{equation}
p(\boldsymbol{\Xi})
= \prod_I\left\{\frac{1}{{\delta_I}\sqrt{2\pi}}
\,\exp
\left(
\frac{{\Xi_I}^2}{2{\delta_I}^2}
\right)
\right\}
\ .
\end{equation}
The exact integral over $\boldsymbol{\Xi}$ would need to be done
numerically for each ${\mu}$, but if $\boldsymbol{\delta}$
are small, one can make the approximation
$e^{{\Lambda_{IJ}}} \approx 1 + {\Lambda_{IJ}}
= 1 + {\Xi_I} + {\Xi_J}$
to convert the likelihood to a Gaussian integral over
$\boldsymbol{\Xi}$ which {\em can} be done analytically.  The result is a
likelihood of the form
\begin{equation}
  p(\mathbf{x}|{\mu})
  = \sqrt{
    \det
    \left(
      \frac
      {
        \mathbf{M}
        \bigl({\mu},\boldsymbol{\sigma},\boldsymbol{\delta}\bigr)
      }{2\pi}
    \right)
  }
  \exp
  \left[
    -\frac{1}{2} \sum_\alpha\sum_\beta ({x_\alpha}-{\mu})
    \,M_{\alpha\beta}
    \bigl({\mu},\boldsymbol{\sigma},\boldsymbol{\delta}\bigr)
    \,
    ({x_\beta}-{\mu})
  \right]
  \ .
\end{equation}
This is the approach which was used for the multi-baseline upper
limits in~\cite{S5LV}.

For the special case of two instruments which make up a single
baseline, the matrix $\mathbf{M}$ reduces to a single number with
value
\begin{equation}
  \label{e:Monepair}
  \mathbf{M}\bigl({\mu},\boldsymbol{\sigma},\boldsymbol{\delta}\bigr)
  \equiv
      \frac{1}
      {{\sigma_{12}^2} + {\mu}^2({\delta_1}^2 + {\delta_2}^2)}\,.
\end{equation}
Comparing \eqref{e:Monepair} to \eqref{e:single}, we see that this
approximation gives the same result as assuming a Gaussian prior in
$\lambda_{12}$ with
${\varepsilon_{12}}^2 ={\delta_1}^2 + {\delta_2}^2$.

\section{Ongoing Work}

More generally, we may be using $\mathbf{x}$ to estimate
multiple physical quantities, 
such as different
spherical harmonic modes of a non-isotropic stochastic GW background
\cite{SPH,S5SPH}.  These methods of analytic marginalization with
either a multivariate Gaussian prior or an approximate likelihood
function can be applied to the effects of calibration uncertainty in
that search as well.  Additionally, these calibration effects may also
be considered in other cross-correlation searches, such as the
modelled cross-correlation search for periodic GW
signals.\cite{cwcrosscorr}

\ack

We wish to thank the organizers of the International Conference on
Gravitation and Cosmology (ICGC 2011, Goa) and our colleagues in the
LIGO Scientific Collaboration and the Virgo Collaboration, especially
Albert Lazzarini.
JTW was supported by NSF grant PHY-0855494.  ELR was supported by the
Max Planck Society.  JDR was supported by NSF grants PHY-0855371 and
HRD-0734800.  EHT was supported by NSF grant PHY-0758035.
This paper has been assigned LIGO Document Number \dcc.

\end{document}